\documentclass[aps,,prl,twocolumn,groupedaddress,showpacs]{revtex4-1}
\usepackage{epsfig}
\usepackage{color}
\usepackage{subfigure}

\begin{document}

\title{Gapless inhomogeneous superfluid phase with spin-dependent disorder}

\author{M.~Jiang$^{1,2}$, R.~Nanguneri,$^{1}$ N.~Trivedi$^3$,
  G.G.~Batrouni,$^{4,5}$ and R.T.~Scalettar$^1$}

\affiliation{$^1$Physics Department, University of California, Davis,
  California 95616, USA}
\affiliation{$^2$Department of Mathematics, University of California, Davis,
  California 95616, USA}
\affiliation{$^3$Department of Physics, The Ohio State University,
  Columbus, OH 43210, USA}
\affiliation{$^4$INLN, Universit\'e de Nice-Sophia Antipolis, CNRS;
  1361 route des Lucioles, 06560 Valbonne, France}
\affiliation{$^5$Institut Universitaire de France}

\begin{abstract}
We show that the presence of a spin-dependent random potential in a
superconductor or a superfluid atomic gas leads to distinct transitions
at which the energy gap and average order parameter vanish, generating
an intermediate gapless superfluid phase, in marked
contrast to the case of spin-symmetric randomness where no such gapless superfluid phase is seen.
By allowing the pairing amplitude to become inhomogeneous, the gapless superconducting phase persists to
considerably higher disorder compared with the prediction of Abrikosov-Gorkov.
The low-lying excited states are located predominantly in regions where the pairing amplitude
vanishes and coexist with the superfluid regions with a finite pairing.
Our results are based on inhomogeneous Bogoliubov-de Gennes mean field theory for a two dimensional attractive Hubbard model
with spin-dependent disorder.
\end{abstract}

\pacs{74.62.En, 37.10.Jk, 74.40.Kb}
\maketitle

\noindent
\underbar{Introduction:}
The interplay of disorder and strong interactions and the mechanisms
underlying metal-insulator transitions is still an open and very
important question.  Traditionally, this problem has been discussed in
the context of solid state systems.  More recently it has become
possible to emulate quantum Hamiltonians using ultracold atomic gases
\cite{cho08}, opening up a new dialog between condensed matter and
atomic physics leading to new insights into the effects of
correlation on randomness.

The Anderson model for localization~\cite{anderson1958}, first proposed
in 1958, describes the possibility of localized electronic states formed
by quantum interference of waves in a random potential. Later, in the
famous ``gang of 4" paper, it was shown using renormalization group
flows of the conductance that a system in two or lower dimensions would
always flow to a localized state for arbitrarily weak disorder.  In
contrast, in three dimensions there is a critical disorder separating
localized and conducting states~\cite{gangof4}.  Over the years the
precise verification of the Anderson model has not been possible because
in real systems interactions cannot be ignored, and, moreover, become
increasingly important as the states get more localized near the
metal-insulator transition.  However, very recently in ultracold atomic
gases the Anderson model has been emulated for the first time by tuning
the interactions to zero~\cite{aspect09} using Feshbach resonance and
the localized wave function has been visualized. In these experiments,
randomness is introduced through optical speckle~\cite{white09} or
bichromatic lattices~\cite{fallani07}.

If attractive interactions between fermions are turned on, such systems become superconducting (for charged electrons) or superfluid (for neutral fermions). These states can exist in two dimensions
in a disordered system below a critical disorder even when all the
single particle states are localized. The ensuing
superconductor-insulator transition has been studied in considerable
detail in the literature~\cite{ghosal98_01, bouadim, dubi08}. A particularly
intriguing aspect of cold atoms is the ability to tune the interactions
either through the Feshbach resonance or through optical lattices.  In
the limit of strong attractive interactions between fermions the problem
crosses over to bosons in a random
potential~\cite{fisher1989,NT,RTS,Prokofeev,Rafael} where new glassy
phases are expected for large disorder.  In this way, the complex
interplay of interactions and disorder is combined with BCS to BEC
crossover physics ~\cite{mohit-NP}.

In this paper, we investigate the effect of a spin-dependent random
potential on a superconductor.  A related problem, that of magnetic
impurities in a superconductor, was studied by Abrikosov and Gorkov~\cite{AG} who
found that the superfluid density was suppressed for fairly small
impurity concentrations.  These diagrammatic calculations were performed
at weak coupling in the BCS regime.  Here, in contrast, we
investigate the intermediate to strongly coupled BCS-BEC crossover regime coupled with
magnetic disorder, a problem that has not been explored so far, and
which is now of very direct experimental relevance to atomic gases.
We show that within Bogliubov-de Gennes mean field
theory spin-dependent disorder exhibits qualitatively different
features from the conventional symmetric case where both species feel
the random potential.  In particular, the gap and order parameter are
driven to zero with increasing disorder, whereas they remain nonzero,
and can even increase, when both species see the same energy landscape.
This destruction of pairing has also been seen in BdG studies in
the presence of spin-independent order and a uniform Zeeman field\cite{dubi08}.

Spin dependent lattices can be achieved, for instance, by utilizing two
atomic hyperfine transitions, D1 and D2 (ground to excited states) of
the fermionic alkali atom $^{40}K$, which can be individually excited by
right- and left-circular polarized light respectively (according to
selection rules). By creating a lattice with independently tunable
components of right- and left-circular polarized standing-wave light, a
spin-dependent lattice is realized~\cite{liu04}. The two fermionic
``spin" species in this case are the two hyperfine states of the
degenerate ground-state manifold split by a small uniform magnetic
field. On the other hand, disorder can be superimposed on the periodic
optical lattices created by interfering counter-propagating lasers by
means fine-grained speckle fields~\cite{white09,aspect09}.
Accordingly, in principle, spin-dependent disorder can be realized by
superimposing on the lattice separate laser beams that have been passed
through a diffuser and focused at the lattice location. A suitable
diffuser would be able to modulate the lattice intensity at each lattice
site in a random way, with the randomness uncorrelated on the scale of
lattice constants~\cite{mandel03,demarco10}.

\noindent
\underbar{Model and Methodology:}
Our starting point is the attractive fermion Hubbard
Hamiltonian with spin dependent disorder,
\begin{eqnarray}
H = &-&t \sum_{\langle {\bf ij}\rangle,\sigma}
\,\,
(c_{{\bf j}\sigma}^{\dagger} c_{{\bf i}\sigma}^{\vphantom{dagger}}
+c_{{\bf i}\sigma}^{\dagger} c_{{\bf j}\sigma}^{\vphantom{dagger}})
\\
&-&|U| \sum_{\bf i} n_{{\bf i}\uparrow} n_{{\bf i}\downarrow}
+ \sum_{\bf i \sigma} \, (\epsilon_{\bf i \sigma}-\mu_\sigma) \,\, n_{{\bf i}\sigma}
\,\,\,
\nonumber
\label{hubham}
\end{eqnarray}
Here $c_{{\bf i}\sigma}^{\dagger} (c_{{\bf
i}\sigma}^{\vphantom{dagger}})$ are fermion creation(destruction)
operators at site ${\bf i}$ for fermionic species (spin or hyperfine
level) $\sigma=\uparrow,\downarrow$.  $t$ is the hopping amplitude
between near neighbor sites $\langle {\bf ij}\rangle$ on a square
lattice.  $t=1$ is the energy unit.  $-|U|$ is an on-site attraction,
and $\mu_\sigma$ is the chemical potential, which in general will depend
on the spin index $\sigma$ in order to maintain equal populations when
the randomness is different for the two species. $\epsilon_{\bf i
\sigma}$ is a local site energy which, in the conventional
``Anderson-Hubbard" Hamiltonian, is chosen to be independent of fermion
species $\sigma$.  Here we focus on the situation where $\epsilon_{{\bf
i} \downarrow}=0$ and $\epsilon_{{\bf i} \uparrow}$ is uniformly
distributed on $[-V,V]$.

In the clean limit, the square lattice attractive Hubbard model is a
supersolid (simultaneous long-range charge density wave and $s$-wave
pair correlations) at $T=0$ and density $\rho=1$ (half-filling).  When
doped to $\rho \neq 1$, the degeneracy is broken, and the attractive
Hubbard Hamiltonian exhibits a finite temperature
Berezinskii-Kosterlitz-Thouless transition to a superconducting phase
with $T_c$ on the order of $0.1 t$~\cite{scalettar89}. In the limit of strong attraction, this
 model crosses over to the Bose-Hubbard
model of tightly bound pairs\cite{fisher89}.

Anderson's theorem provides a first insight into the effect of
randomness on the clean attractive Hubbard model, suggesting that, at
least for weak disorder, superconductivity will survive.  A number of
numerical studies have quantitatively explored this model
\cite{trivedi96}.  Here we will focus our comparison on the
Bogoliubov-de Gennes (BdG) treatment of Ghosal {\it et
al.}\cite{ghosal98_01} since that is also the methodology employed here.

The interaction term in $H$ can be decoupled in different (charge, pairing, spin) channels.  Since
$U<0$ we focus on pairing and write,
\begin{eqnarray}
H_{\rm eff} = &-&t \sum_{\langle {\bf ij} \rangle,\sigma}
(c^{\dagger}_{{\bf i}\sigma}c^{\phantom{\dagger}}_{{\bf j}\sigma} +
c^{\dagger}_{{\bf j}\sigma}c^{\phantom{\dagger}}_{{\bf i}\sigma} ) +
\sum\limits_{i\sigma} (\epsilon_{i\sigma}-\widetilde{\mu}_{{\bf i}\sigma}) n_{i\sigma} \nonumber
\\ &+& \sum_{\bf i} [\Delta^{\phantom{\dagger}}_{\bf i}
c^{\dagger}_{{\bf i}\uparrow}c^{\dagger}_{{\bf i}\downarrow}
+\Delta^{*}_{\bf i} c^{\phantom{\dagger}}_{{\bf i}\downarrow}
c^{\phantom{\dagger}}_{{\bf i}\uparrow} ] \,\,
\end{eqnarray}
where $\widetilde{\mu}_{{\bf i}\sigma}=\mu_{\sigma} + |U|\langle
n_{i,-\sigma} \rangle$. $H_{\rm eff}$ is diagonalized via the
Bogoliubov transformation,
\begin{eqnarray}
c^{\phantom{\dagger}}_{{\bf i}\uparrow} &=&
\sum_{n} [ \, \gamma^{\phantom{\dagger}}_{n\uparrow} u_{{\bf i}n} -
\gamma^{\dagger}_{n\downarrow} v^{*}_{{\bf i}n}  ]
\nonumber \\
c^{\phantom{\dagger}}_{{\bf i}\downarrow} &=&
\sum_{n} [ \, \gamma^{\phantom{\dagger}}_{n\downarrow} u_{{\bf i}n} +
\gamma^{\dagger}_{n\uparrow} v^{*}_{{\bf i}n}] \,\,
\end{eqnarray}
In the clean system the eigenfunctions $u_{{\bf i}n}$ and $v_{{\bf
    i}n}$ are plane waves.  In the presence of disorder they are
obtained by numerical diagonalization.  The local order parameter and
density are determined self-consistently,
\begin{eqnarray}
\Delta_{\bf i}&=&-|U|\langle c^{\phantom{\dagger}}_{{\bf i}\downarrow}
c^{\phantom{\dagger}}_{{\bf i}\uparrow} \rangle
= -|U| \sum_n f(E_n)  u_{{\bf i}n} v^{*}_{{\bf i}n}
\\
\langle n^{\phantom{\dagger}}_{{\bf i}\uparrow} \rangle &=& \sum_n
f(E_n) | u^{\phantom{\dagger}}_{{\bf i}n}|^2
\hskip0.30in
\langle n^{\phantom{\dagger}}_{{\bf i}\downarrow} \rangle = \sum_n
f(-E_n) | v^{\phantom{\dagger}}_{{\bf i}n}|^2
\nonumber
\end{eqnarray}
(where $f$ is the Fermi function), as are the chemical potentials
required to achieve the desired density\cite{ghosal98_01}.  These
self-consistency conditions are equivalent to minimizing the free
energy.  The correctness and efficiency of our codes have been checked
by using different initial configurations of particle density and local
pairing amplitude and checking for convergence. As expected, we found more iterations are needed
for convergence near phase transitions. We define a spatially
averaged order parameter, $\Delta_{\rm op}$, from $\Delta_{\bf i}$.  The
BdG spectrum can be used to determine the energy gap, $E_{\rm gap}$,
which is the lowest eigenvalue above the chemical potential.  The
spectrum and eigenfunctions also detemine the density of states.  Note
that unlike the spin-independent case the eigenvalues do not come in
$\pm$ pairs and the distances of closest eigenvalues below and above the
chemical potential are in general different.

The key conclusions of the BdG treatment of Ghosal {\it et al.}
\cite{ghosal98_01} for the usual spin-symmetric case $\epsilon_{{\bf
i}\uparrow}= \epsilon_{{\bf i}\downarrow}$ are as follows: As disorder
is increased the energy gap and lattice-averaged order parameter
decrease, but never go to zero.  In particular, even though the
distribution of local pairing amplitudes has significant weight near
zero, a finite spectral gap persists owing to remnant superconducting
islands and, importantly, significant overlap of the low energy excited
states with these islands.
Anderson's theorem concerning the survival of superconductivity is obtained under two assumptions: pairing of exact eigenstates
and a further assumption that the kernel in the gap equation is spatially uniform. It is possible to generalize the calculation within
pairing of exact eigenstates to allow for spatial structure in the pairing amplitude. BdG calculations go beyond pairing of exact eigenstates
and allow for a full treatment of the amplitude fluctuations in response to an underlying random potential.
In order to see either the finite temperature phase transition from a superfluid to a non-superfluid state or the quantum phase transition from a superfluid to an insulator, thermal and quantum phase fluctuations must be included in the inhomogeneous BdG state.

We will see that even at the level of pairing of exact eigenstates or BdG the
situation is dramatically transformed by spin-dependent disorder.

\begin{figure}
\centerline{\epsfig{figure=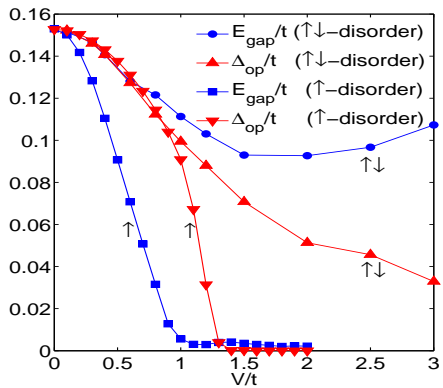,height=5.5cm,width=6.5cm,angle=0,clip}}
\vskip -0.2cm
\caption{The lattice-averaged order parameter $\Delta_{\rm op}$ and
  the energy gap $E_{\rm gap}$ are shown as functions of the disorder
  strength $V$.  In the spin-symmetric case considered in
  [\onlinecite{ghosal98_01}], these quantities never go to zero even
  at large $V$.  In contrast, when the disorder is applied only to one
  spin species, sharp transitions are observed.  The small ``tail'' in
  $E_{\rm gap}$ goes to zero as the lattice size increases. $N=24\times24, \langle n \rangle=0.875, U/t=-1.5$.}
\label{Egap_OP_V_U15}
\end{figure}

\begin{figure}
\psfig{figure=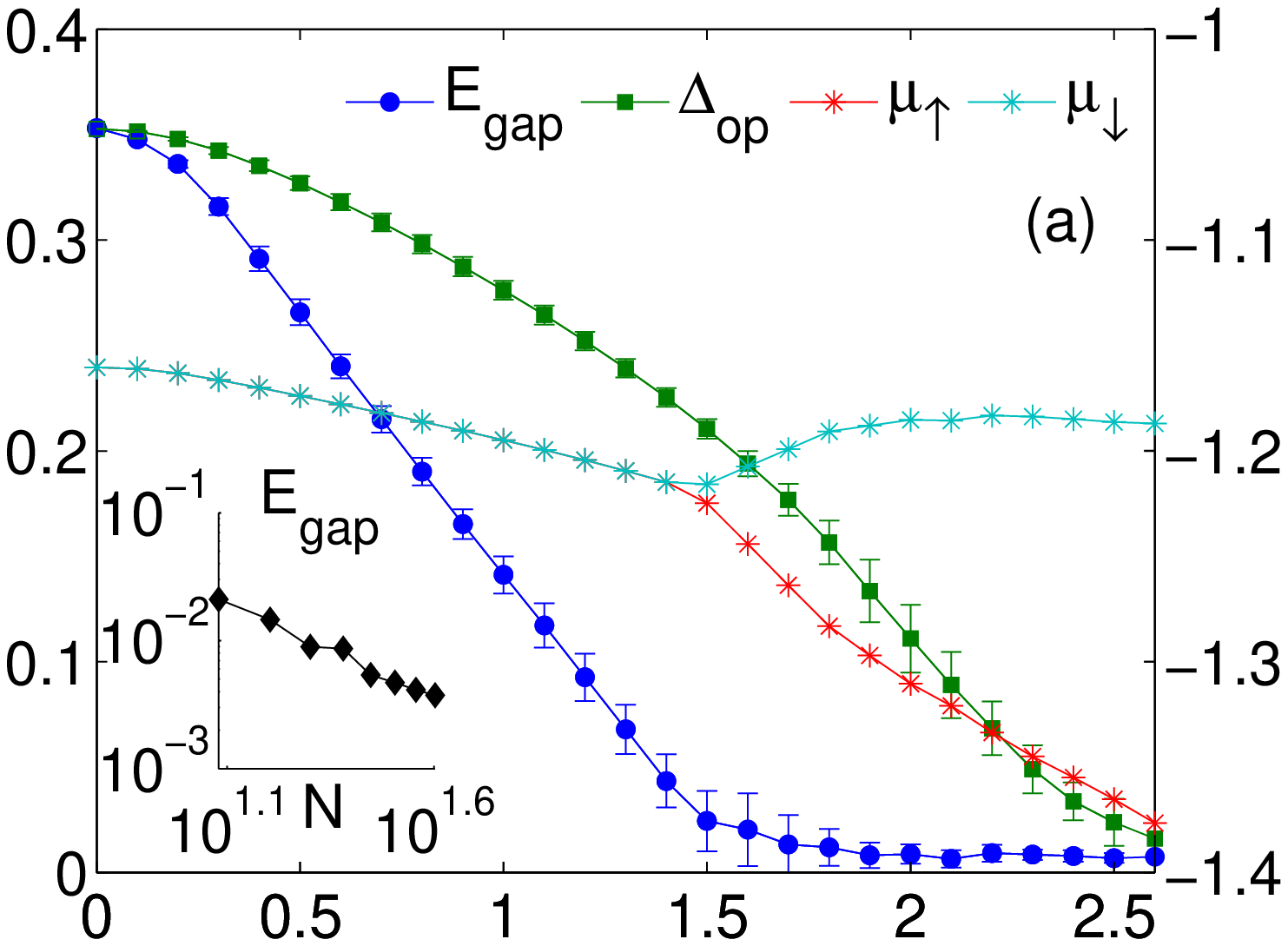,height=4.6cm,width=7.0cm,angle=0,clip}\\
\vskip -0.2cm
\epsfig{figure=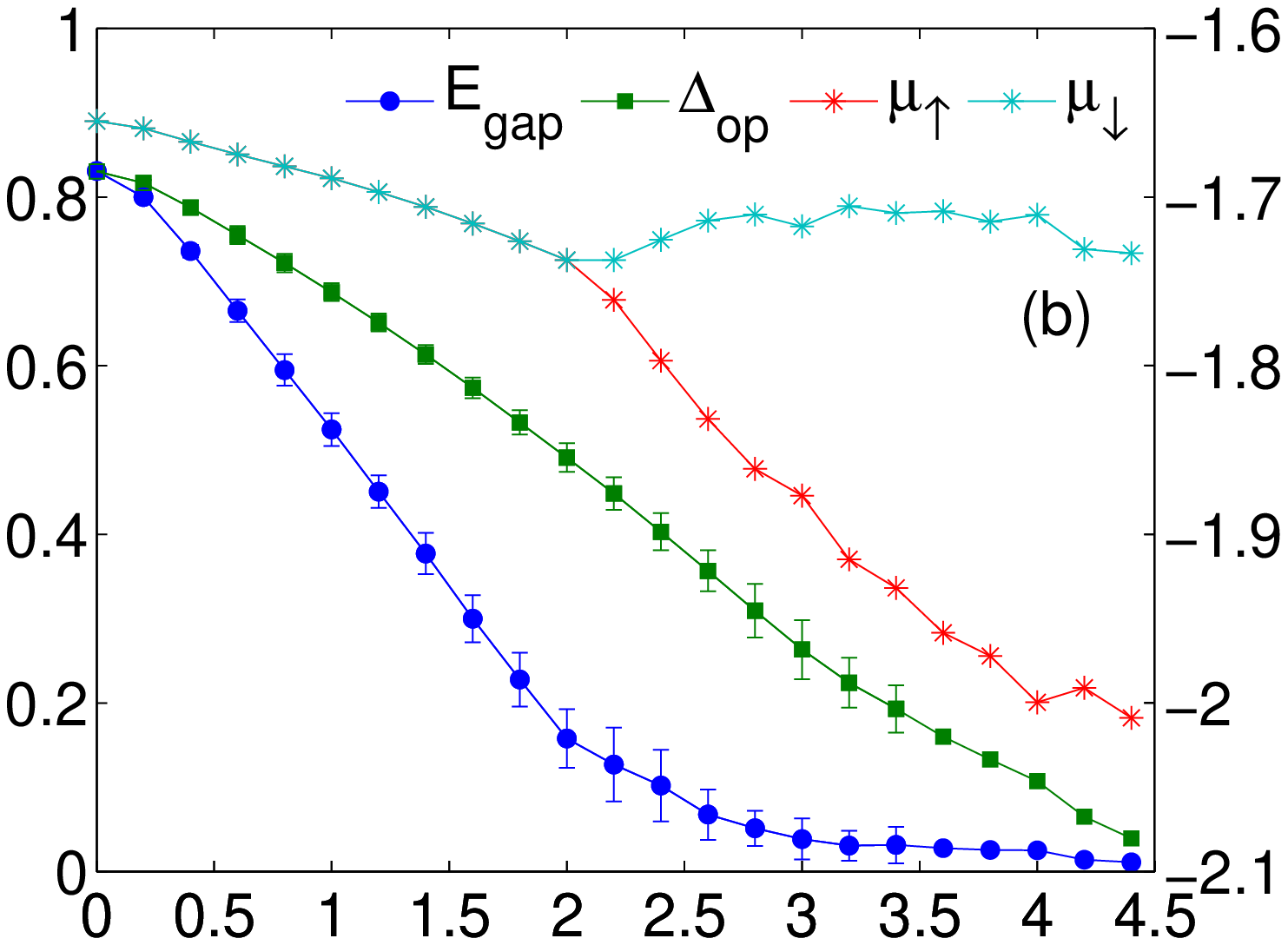,height=4.6cm,width=7.0cm,angle=0,clip}\\
\vskip -0.2cm
\psfig{figure=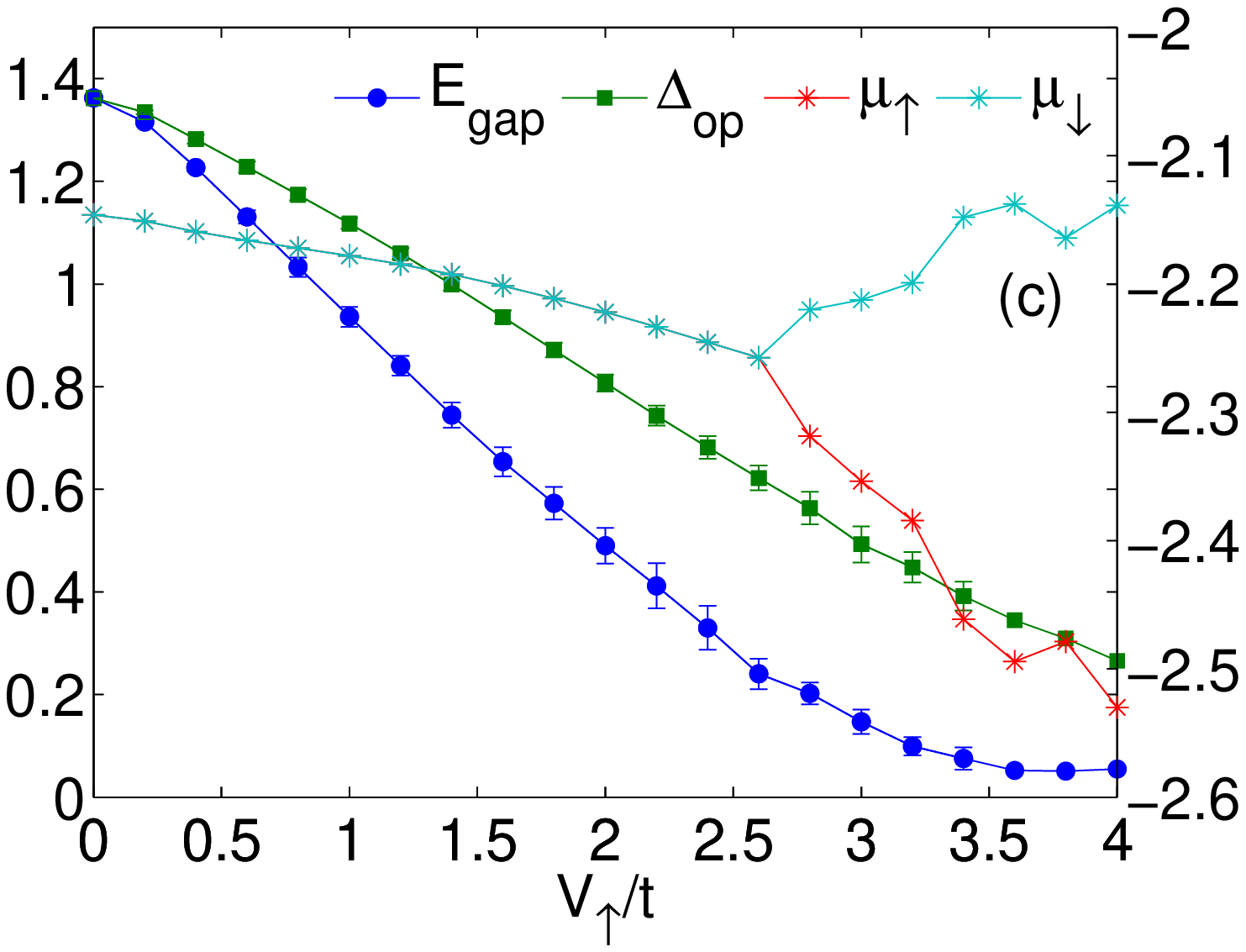,height=4.6cm,width=7.0cm,angle=0,clip}\\
\vskip -0.12cm
\psfig{figure=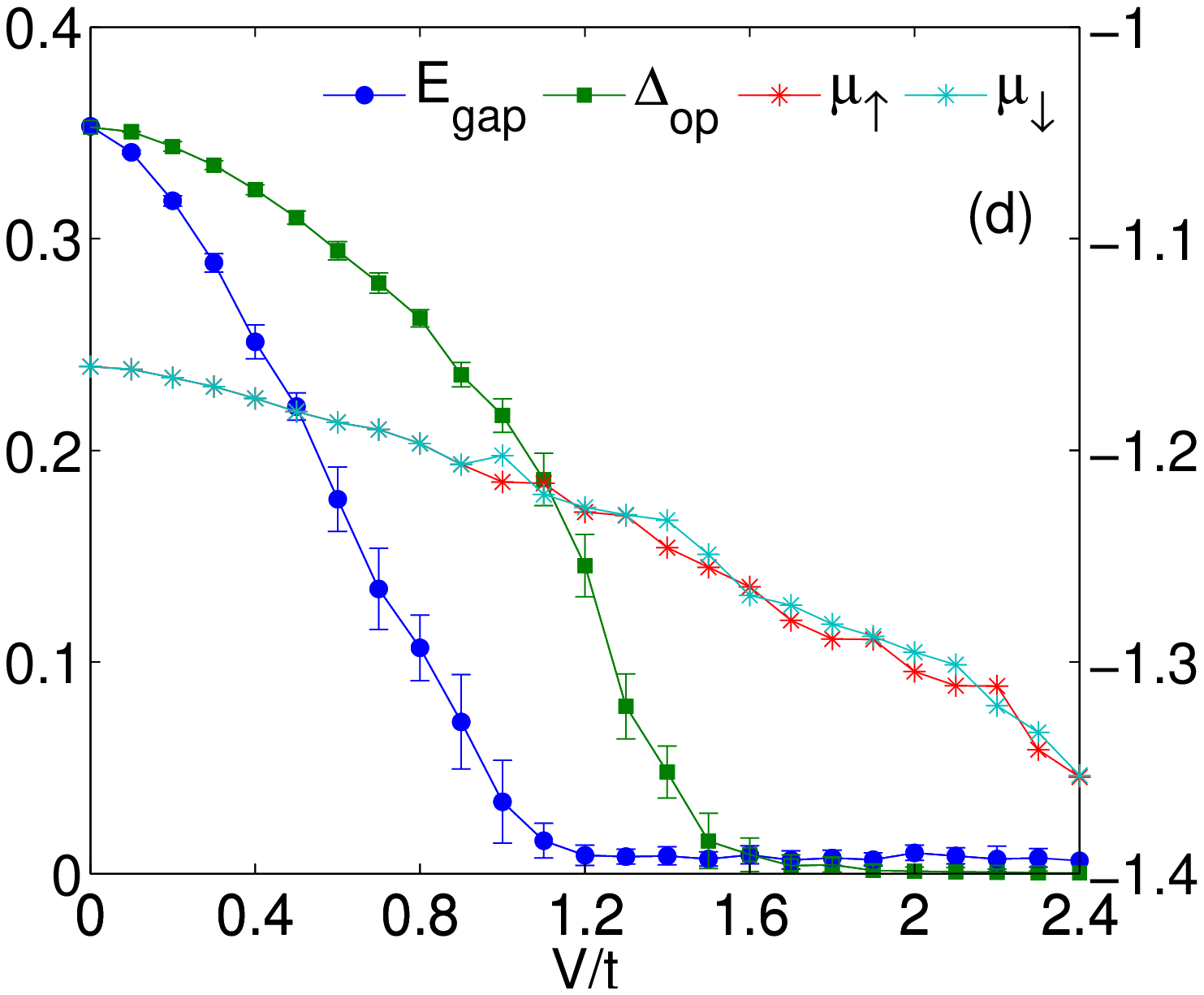,height=4.6cm,width=7.0cm,angle=0,clip}
\begin{caption}
{The energy gap $E_{\rm gap}$, average order parameter
$\Delta_{\rm op}$ (left $y$-axis) and chemical potentials $\mu_{\sigma}$
(right $y$-axis) are shown as functions of disorder strength $V$.
Panels (a,b,c) show results for the case when $-V< \epsilon_{{\bf
    i}\uparrow}<V$ and $\epsilon_{{\bf i}\downarrow}=0$ for three
values, $U/t = -2, -3, -4$. Panel (d) shows a case when both
$\epsilon_{{\bf i}\sigma}$ are uniformly distributed on $[-V,V]$ but
are chosen independently, for $U/t = -2$. The qualitative behavior for
these two types of spin-asymmetric disorder is similar: There are two
transitions, first from gapped to gapless superfluid and then to a
phase in which $\Delta_{\rm op}=0$ as well.  As expected,
$\mu_{\sigma}$ is spin-independent in the situation of panel (d) when
both species see disorder of the same overall strength (although
distinct realizations). $N=24\times24, \langle n\rangle=0.875$. The
results are averaged over 10 realizations of randomness.}
\end{caption}
\label{gapopmun0875U15N24alpha1}
\end{figure}


\noindent
\underbar{Results:}
Figure \ref{Egap_OP_V_U15} shows the evolution of the energy gap $E_{\rm
gap}$ and order parameter $\Delta_{\rm op}$ with increasing randomness
$V$ for density $\langle n \rangle=0.875$ and on-site attraction $U=-1.5t$.  In the
spin-symmetric case, confirming the results of
[\onlinecite{ghosal98_01}], we find $E_{\rm gap}$ and $\Delta_{\rm op}$
do not vanish.  However, in the spin-asymmetric case, when
$\epsilon_{{\bf i}\downarrow}=0$, there are instead sharp transitions
for both quantities.  Interestingly, $E_{\rm gap}$ and $\Delta_{\rm op}$
do not vanish simultaneously, but instead three phases are present: a
superconductor characterized by $E_{\rm gap}$ and $\Delta_{\rm op}$ both
nonzero at small disorder $V$, a gapless
superconductor\cite{deGennes,AG,Lamacraft} at intermediate $V$, and a
$E_{\rm gap}=\Delta_{\rm op}=0$ phase at large $V$.

The gapless superfluid has also been observed in other situations in
which spin symmetry is broken, {\it e.g.}~the mismatched Fermi surface
considered in [\onlinecite{feiguin09}].  In such situations, paired
regions of the Fermi sea coexist with regions occupied by only a single
spin species.  These unpaired pockets lead to a gapless superfluid phase
between the normal and BCS regimes, much as occurs in
Fig.~\ref{Egap_OP_V_U15}.  Indeed, as we shall demonstrate later, in the
attractive Anderson Hubbard Hamiltonian considered here, the real space
distribution of $\Delta_{\bf i}$, indicates coexisting {\it real space}
domains which are in direct analogy to the coexisting {\it momentum
space} domains of [\onlinecite{feiguin09}] and related work.

\begin{figure}
\vskip -8.0cm
\psfig{figure=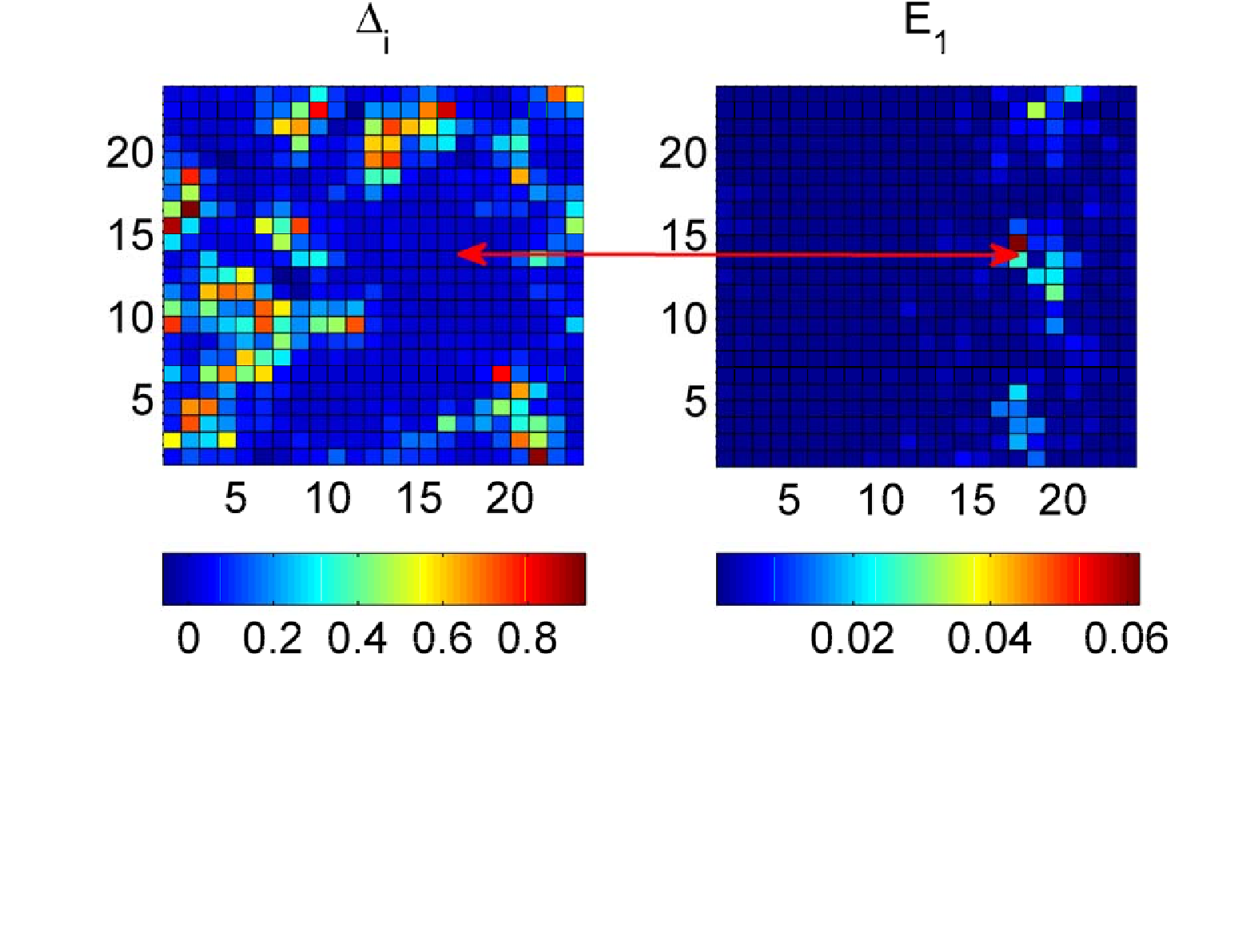,height=14.0cm,width=11.0cm,angle=0,clip}
\vskip -2.0cm
\begin{caption}
{ Left panel: The distribution of the local order parameter
  $\Delta_{\bf i}$.  Right panel: The first excited state wave
  function.  Here $U=-3t$ and $V_{\uparrow}=4t$.  The average order
  parameter is non-zero, but because the first excited state has
  weight in the region where the local values are zero there is no
  energy gap.  Hence the system is a gapless superfluid.  }
\end{caption}
\label{domains}
\end{figure}

Figures 2(abc) provide further details for the evolution of $E_{\rm
gap}$ and $\Delta_{\rm op}$ for three interaction strengths, $U/t= -2.0,
-3.0$ and $-4.0$.  The transition points move to larger disorder
strength, as expected, as the attractive interaction is increased.  At
all $U/t$, there is a residual nonzero value of $E_{\rm gap}$ which
scales to zero with increasing system size.  In fact, the transition
point from gapped to gapless superconductor is best signalled by a
bifurcation of the chemical potentials $\mu_{\uparrow} \neq
\mu_{\downarrow}$ required to maintain a spin-balanced population
$\rho_{\uparrow}=\rho_{\downarrow}$.

In order to determine whether such transitions are generic to the
breaking of spin symmetry in the disorder we consider, we show in
Fig.~2d a case where both species see disorder drawn from the same
distribution $[-V,V]$ but with different realizations.  That is, spin
$\uparrow$ and spin $\downarrow$ fermions see different disordered
landscapes.  The basic features are the same as the case when only one
species sees the disorder (Figs.~2abc): there are two separate
transitions at which $E_{\rm gap} \rightarrow 0$ and $\Delta_{\rm op}
\rightarrow 0$.  As expected, the chemical potentials do not bifurcate
in the case of Fig.~2d.  Nevertheless, there is still a signature of the
vanishing of $E_{\rm gap}$ in the scatter of the chemical potential
data.

To understand the physics of the gapless superfluid we show, in Fig.~3,
the spatial distribution of the local order parameter $\Delta_{\bf i}$
(left) and the first excited state wave function $\Psi^*_{\bf
i}$(right).  The system is highly inhomogeneous and, although many sites
have nonzero $\Delta_{\bf i}$, so that the average order parameter is
substantial, there are also large domains where the local order
parameter vanishes.  The first excited state lives predominantly on the
latter set of sites, leading to a vanishing energy gap.  In the gapped
superfluid phase at weaker disorder, $\Psi^*$ substantially overlaps
regions of non-zero order parameter.  Similarly, in the canonical case
of spin-independent disorder the sites at which the first excited state
has non-zero amplitude {\it coincide} with those of non-zero order
parameter.

\begin{figure}
\psfig{figure=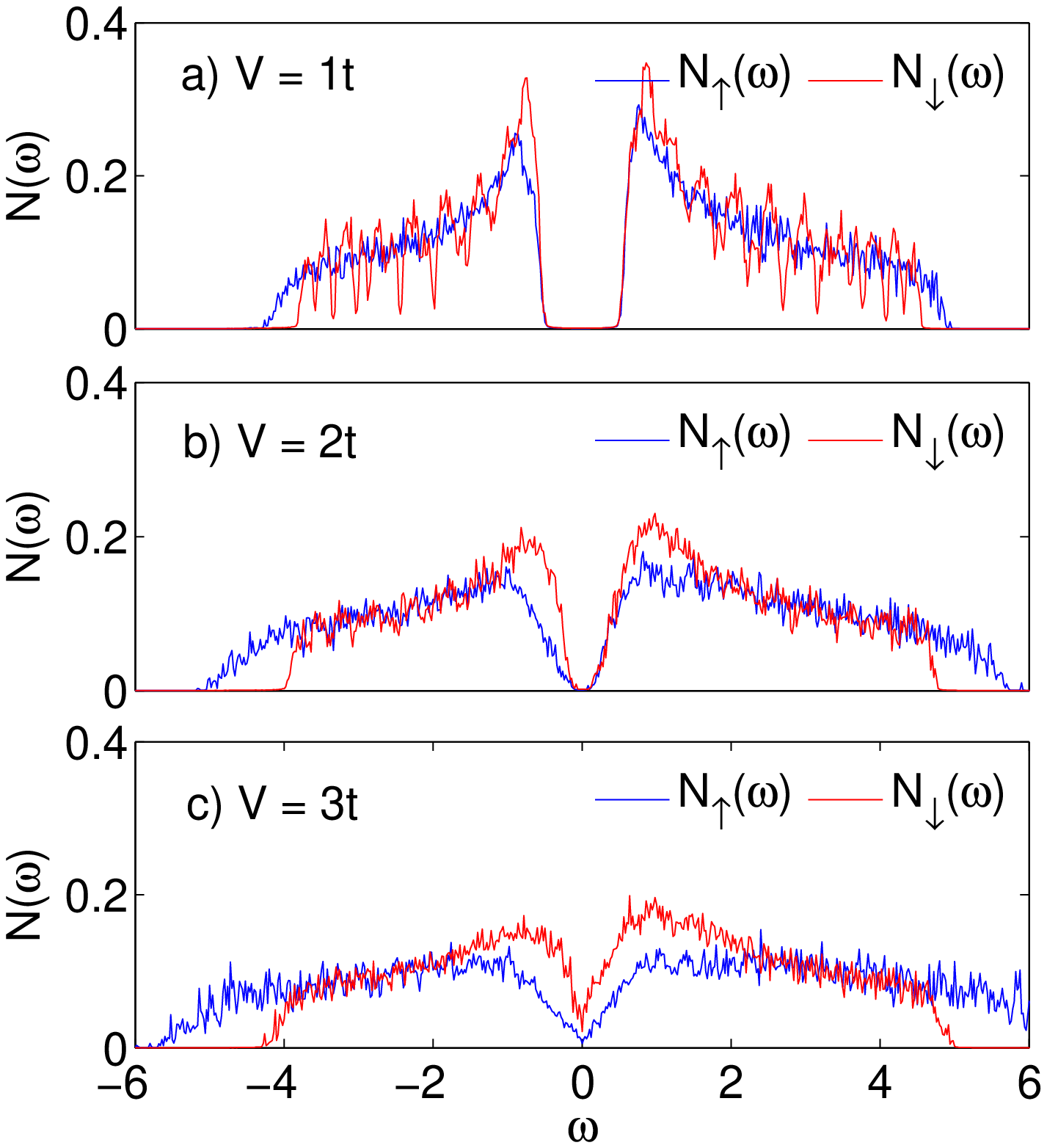,height=6.5cm,width=7.0cm,angle=0,clip}
\begin{caption}
{Density of states at $U=-3t$ for (top to bottom) $V_\uparrow = 1.0t,
  2.0t$ and $3.0t$.  The gap closes for $V_\uparrow = 3.0t$.  See also
  Fig.~2(b).}
\end{caption}
\label{dosU}
\end{figure}

We can also infer the phase from the density of states $N(\omega)$
obtained as a histogram of the BdG eigenvalues.  Note again that the
breaking of spin symmetry destroys the usual appearance of eigenvalues
in $\pm E_n$ pairs.  Figure 4 shows the density of states
\begin{eqnarray}
N_{\uparrow}(\omega) = \frac{1}{N} \sum\limits_{n,{\bf i}} |u_{{\bf i}n}|^{2} \delta (\omega-E_{n})
\nonumber \\
N_{\downarrow}(\omega) = \frac{1}{N} \sum\limits_{n,{\bf i}} |v_{{\bf i}n}|^{2} \delta (\omega+E_{n})
\end{eqnarray}
for $U=-3t$ and a range of disorder strengths.  The gap is nonzero for
$V_\uparrow=1.0t$ but has closed by the time $V_\uparrow = 3.0t$, in
agreement with Fig.~2(b).

\begin{figure}
\psfig{figure=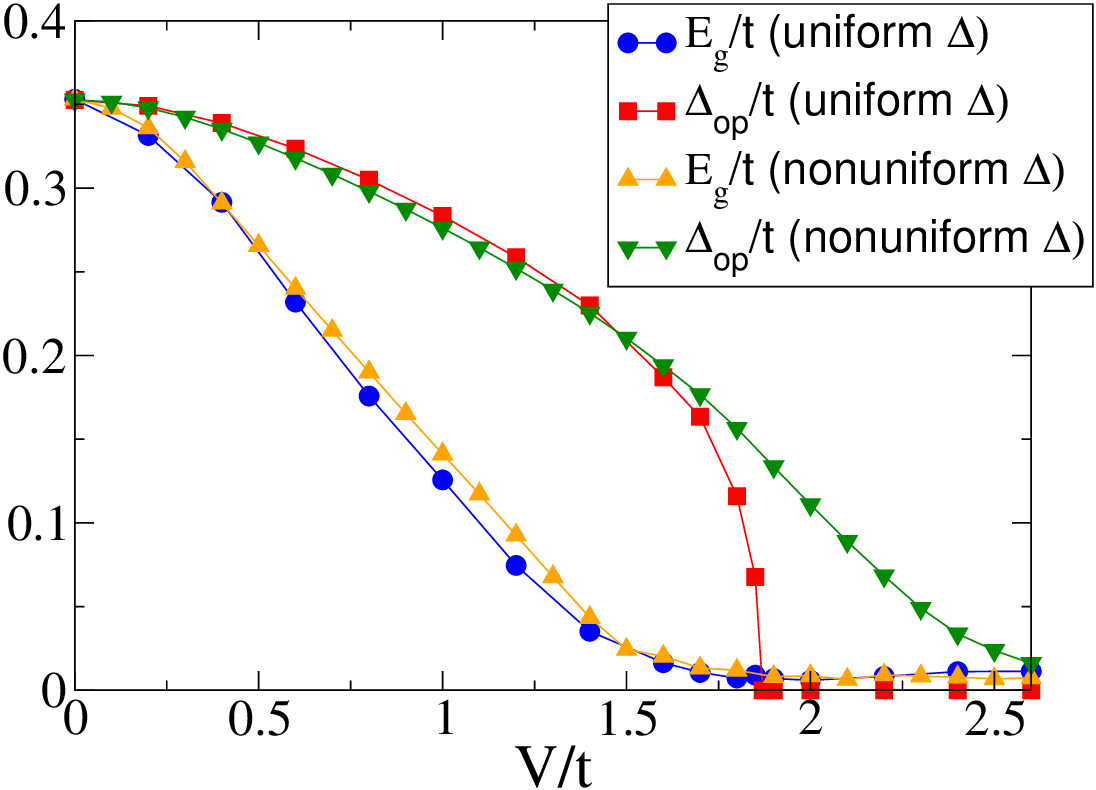,height=5.0cm,width=7.0cm,angle=0,clip}
\begin{caption}
{The behavior of the energy gap and the order parameter for the same disorder strength for the case
when the local order parameter is forced to be uniform to conform to the conditions for which the classic Abrikosov-Gorkov theory~\cite{AG}
was developed and its comparison with a spatially non-uniform pairing amplitude. Notice that a finite order parameter can exist up to much higher disorder
when the pairing amplitude is allowed to be non-uniform. $N=24\times24, U/t=-2, \langle n \rangle=0.875$.}
\end{caption}
\label{AGcomparison}
\end{figure}

\noindent
\underbar{Discussion:}The effects of spin-dependent disorder on the superconducting phase in
the attractive Hubbard Hamiltonian differ fundamentally from
conventional chemical potential disorder.  Sharp transitions at which
the energy gap and average order parameter vanish can clearly be seen as
the randomness increases.  This can be attributed at the fundamental
level to the breaking of time reversal symmetry, which can arise in the
solid state by the spin-dependent scattering produced by magnetic
impurities.  In fact, the spin-dependent site energy $\epsilon_{{\bf
i}\uparrow} n_{{\bf i}\uparrow} +\epsilon_{{\bf i}\downarrow} n_{{\bf
i}\downarrow}$ in Eq.~\ref{hubham}, can be thought of as a combination
of a local chemical potential $ \frac{1}{2} (\epsilon_{{\bf i}\uparrow}
+ \epsilon_{{\bf i}\downarrow} ) \, (n_{{\bf i}\uparrow} + n_{{\bf
i}\downarrow}) $ and a local Zeeman field, $\frac{1}{2} (\epsilon_{{\bf
i}\uparrow} - \epsilon_{{\bf i}\downarrow} ) \, (n_{{\bf i}\uparrow} -
n_{{\bf i}\downarrow})$.  The Zeeman term breaks time reversal symmetry,
providing one way to interpret the results presented here. The occurrence of sharp transitions can also be understood in analogy with
exotic paired states in other situations like Fulde-Ferrell-Larkin-
Ovchinnikov\cite{fulde64,larkin64} with mismatched Fermi
surfaces\cite{feiguin09}.  In these systems, electrons of one species
with a particular {\it momentum} cannot find partners, and hence a
superfluid and unpaired electrons coexist.  An analogous phenomenon
occurs in {\it real space} in the Hamiltonian considered in this paper:
the spin-dependent randomness provided regions of the lattice where
fermions cannot find partners, while in other regions pairing can
continue.

\noindent {\underbar{Comparison with Abrikosov-Gorkov theory:}}
It is useful to compare our results in Figs.2 and 3 against the early work by Abrikosov and Gorkov (AG)~\cite{AG} where they first predicted the possibility of gapless
superconductivity for spin-dependent scattering, a phase with finite order parameter but no gap. Their result was obtained within perturbation theory in a regime of weak disorder
assuming that the local pairing amplitude is homogeneous.

In Fig.5 we compare our results for the energy gap and order parameter obtained from the inhomogeneous BdG equations as discussed above against a restricted calculation
where we enforce a spatially uniform local pairing amplitude $\bar{\Delta}\equiv(1/N) \sum_i\Delta({\bf r}_i)$ by finding a self consistent solution for the average pairing amplitude $\bar\Delta$. The significant aspects of these results are:
(1) When the local pairing amplitude is forced to be uniform, there is a small region in disorder strength where a gapless phase exists, in agreement with AG theory~\cite{AG}.
(2) When the local pairing amplitude is allowed to vary spatially, a much larger region of gapless superconductivity opens up. The gap is not affected much by the local pairing amplitude variations but the disorder strength up to which
a finite order parameter exists as deduced from long range superconducting correlations is greatly enhanced.

In that respect the gapless superconducting phase that is reported here is different in nature from that in the earlier work because inhomogeneity of the local
order parameter plays a vital role in enhancing the order parameter.

\noindent
\underbar{Conclusion:}
A key goal of optical lattice emulation is to observe Quantum Phase
Transitions (QPTs) in ultra-cold atomic gases, analogous to those
believed to underlie condensed matter phenomena such as high temperature
superconductivity.  Our work suggests that, for exploring QPTs
associated with the interplay of attractive interactions and disorder,
the sorts of spin-dependent lattices achieved experimentally
\cite{demarco10,mandel03,feiguin09} might be especially useful.

Several fundamental conceptual questions remain to be addressed by these
experiments, and by more precise theoretical approaches such as Quantum
Monte Carlo.  One is the construction of a unified phase diagram in
which spin-dependence arising from the scattering, spin dependent
disorder considered here, and from imbalanced populations, FFLO
physics, are both included.  Specific issues include the conditions for
the existence of gapless superfluidity and whether an inhomogeneous
order parameter is essential for its formation.  Another outstanding
question concerns transport properties: Starting from the weakly
interacting limit, in which one species feels disorder and the other
does not, the system is `half-metal' with distinct localized and
itinerant components.  As an attractive interaction $U$ is turned on,
the delocalized particles will feel an {\it induced} randomness as they
interact with the localized species.  Is there a critical $U$ separating
distinct half-metallic phases from phases in which the two components
have identical transport characteristics?

We acknowledge financial support from ARO Award W911NF0710576 with
funds from the DARPA OLE Program, the CNRS-UC Davis EPOCAL LIA joint
research grant, the Department of Energy under
DE-FG52-09NA29464, and useful input from B. Brummels.

\end{document}